\documentclass[12pt]{article}

\usepackage{mathrsfs}
\usepackage[T1]{fontenc}
\usepackage{mathpazo}
\usepackage{setspace}
\usepackage{amsfonts}
\usepackage{amssymb}
\usepackage{amsmath}
\usepackage{esint}                   
\usepackage{epsfig}
\usepackage{latexsym}
\usepackage{color}
\usepackage{graphicx}
\usepackage{hyperref}
\usepackage{nicefrac}
\usepackage[latin1]{inputenc}
\usepackage{pstricks}
\usepackage{slashed}
\usepackage{multirow}
\usepackage{ulem}
\usepackage{comment}
\usepackage[nosort]{cite}
\usepackage{datetime}
\usepackage{tikz-cd}
\usepackage{float}

\def\hybrid{\topmargin -30pt    \oddsidemargin 0pt 
        \headheight 0pt \headsep 0pt
        \textwidth 6.25in       
        \textheight 9.5in       
        \marginparwidth .875in
        \parskip 5pt plus 1pt   \jot = 1.5ex}

\hybrid

\def\baselinestretch{1.2}

\catcode`\@=11

\def\marginnote#1{}
\newcount\hour
\newcount\minute
\newtoks\amorpm
\hour=\time\divide\hour by60
\minute=\time{\multiply\hour by60 \global\advance\minute by-\hour}
\edef\standardtime{{\ifnum\hour<12 \global\amorpm={am}%
        \else\global\amorpm={pm}\advance\hour by-12 \fi
        \ifnum\hour=0 \hour=12 \fi
        \number\hour:\ifnum\minute<10 0\fi\number\minute\the\amorpm}}
\edef\militarytime{\number\hour:\ifnum\minute<10 0\fi\number\minute}

\def\draftlabel#1{{\@bsphack\if@filesw {\let\thepage\relax
   \xdef\@gtempa{\write\@auxout{\string
      \newlabel{#1}{{\@currentlabel}{\thepage}}}}}\@gtempa
   \if@nobreak \ifvmode\nobreak\fi\fi\fi\@esphack}
        \gdef\@eqnlabel{#1}}
\def\@eqnlabel{}
\def\@vacuum{}
\def\draftmarginnote#1{\marginpar{\raggedright\scriptsize\tt#1}}

\def\draft{\oddsidemargin -.5truein
        \def\@oddfoot{\sl preliminary draft \hfil
        \rm\thepage\hfil\sl\today\quad\militarytime}
        \let\@evenfoot\@oddfoot \overfullrule 3pt
        \let\label=\draftlabel
        \let\marginnote=\draftmarginnote
   \def\@eqnnum{(\theequation)\rlap{\kern\marginparsep\tt\@eqnlabel}%
\global\let\@eqnlabel\@vacuum}  }

\def\draft2{
        \def\@oddfoot{\sl preliminary draft \hfil
        \rm\thepage\hfil\sl\today\quad\militarytime}
        \let\@evenfoot\@oddfoot \overfullrule 3pt
        \let\marginnote=\draftmarginnote
   \def\@eqnnum{(\theequation)\rlap{\kern\marginparsep\tt\@eqnlabel}%
\global\let\@eqnlabel\@vacuum}  }

\def\preprint{\twocolumn\sloppy\flushbottom\parindent 2em
        \leftmargini 2em\leftmarginv .5em\leftmarginvi .5em
        \oddsidemargin -.5in    \evensidemargin -.5in
        \columnsep .4in \footheight 0pt
        \textwidth 10.in        \topmargin  -.4in
        \headheight 12pt \topskip .4in
        \textheight 6.9in \footskip 0pt
        \def\@oddhead{\thepage\hfil\addtocounter{page}{1}\thepage}
        \let\@evenhead\@oddhead \def\@oddfoot{} \def\@evenfoot{} }

\def\numberbysection{\@addtoreset{equation}{section}
        \def\theequation{\thesection.\arabic{equation}}}

\def\underline#1{\relax\ifmmode\@@underline#1\else
        $\@@underline{\hbox{#1}}$\relax\fi}

\def\titlepage{\@restonecolfalse\if@twocolumn\@restonecoltrue\onecolumn
     \else \newpage \fi \thispagestyle{empty}\c@page\z@
        \def\thefootnote{\fnsymbol{footnote}} }

\def\endtitlepage{\if@restonecol\twocolumn \else \newpage \fi
        \def\thefootnote{\arabic{footnote}}
        \setcounter{footnote}{0}}  

\catcode`@=12
\relax

\def\figcap{\section*{Figure Captions\markboth
        {FIGURECAPTIONS}{FIGURECAPTIONS}}\list
        {Figure \arabic{enumi}:\hfill}{\settowidth\labelwidth{Figure
999:}
        \leftmargin\labelwidth
        \advance\leftmargin\labelsep\usecounter{enumi}}}
 \relax
\def\tablecap{\section*{Table Captions\markboth
        {TABLECAPTIONS}{TABLECAPTIONS}}\list
        {Table \arabic{enumi}:\hfill}{\settowidth\labelwidth{Table
999:}
        \leftmargin\labelwidth
        \advance\leftmargin\labelsep\usecounter{enumi}}}
 \relax
\def\reflist{\section*{References\markboth
        {REFLIST}{REFLIST}}\list
        {[\arabic{enumi}]\hfill}{\settowidth\labelwidth{[999]}
        \leftmargin\labelwidth
        \advance\leftmargin\labelsep\usecounter{enumi}}}
 \relax
\makeatletter
\newcounter{pubctr}
\def\publist{\@ifnextchar[{\@publist}{\@@publist}}
\def\@publist[#1]{\list
        {[\arabic{pubctr}]\hfill}{\settowidth\labelwidth{[999]}
        \leftmargin\labelwidth
        \advance\leftmargin\labelsep
        \@nmbrlisttrue\def\@listctr{pubctr}
        \setcounter{pubctr}{#1}\addtocounter{pubctr}{-1}}}
\def\@@publist{\list
        {[\arabic{pubctr}]\hfill}{\settowidth\labelwidth{[999]}
        \leftmargin\labelwidth
        \advance\leftmargin\labelsep
        \@nmbrlisttrue\def\@listctr{pubctr}}}
 \relax
\makeatother

\def\be{\begin{equation}}
\def\ee{\end{equation}}
\def\ba{\begin{eqnarray}}
\def\ea{\end{eqnarray}}

\def\del{\partial}

\def\k{\kappa}
\def\r{\rho}
\def\a{\alpha}

\def\b{\beta}

\def\g{\gamma}
\def\G{\Gamma}
\def\d{\delta}
\def\D{\Delta}

\def\th{\theta}

\def\m{\mu}
\def\n{\nu}

\def\l{\lambda}
\def\L{\Lambda}
\def\s{\sigma}

\def\cL{{\cal L}}

\def\no{\noindent}

\def\qq{\qquad}

\def\IR{\relax{\rm I\kern-.18em R}}

\def\inv{^{\raise.0ex\hbox{${\scriptscriptstyle -}$}\kern-.05em 1}}

\def \z { {\bar z} }

\def \ha {{\frac{1}{2}}}

\def \ov {\over}

\def\tr{\textrm{Tr}}

\begin{document}


\renewcommand{\theequation}{\thesection.\arabic{equation}}
\csname @addtoreset\endcsname{equation}{section}

\begin{titlepage}
\begin{center}

\renewcommand*{\thefootnote}{\arabic{footnote}}

\phantom{xx}
\vskip 0.5in

{\large {\bf Novel integrable interpolations}}

\vskip 0.5in

{\bf Georgios Itsios},\footnote{E-mail:~gitsios@phys.uoa.gr}\hskip .2cm
{\bf Konstantinos Sfetsos}\footnote{E-mail:~ksfetsos@phys.uoa.gr}\hskip .15cm and \ 
{\bf Konstantinos Siampos}\footnote{E-mail:~konstantinos.siampos@phys.uoa.gr}

\vskip 0.1in

Department of Nuclear and Particle Physics, \\
Faculty of Physics, National and Kapodistrian University of Athens, \\
Athens 15784, Greece \\
\vskip .3 cm


\vskip .2in

\end{center}

\vskip .4in

\centerline{\bf Abstract}

\no
A novel class of integrable $\sigma$-models interpolating between exact coset conformal field theories in the IR and hyperbolic spaces in the UV is constructed. 
We demonstrate the relation to the asymptotic limit of $\lambda$-deformed models  for cosets of non-compact groups. An integrable model interpolating between 
two spacetimes with cosmological and black hole interpretations and exact conformal field theory descriptions is also provided. In the process of our work, a 
new zoom-in limit, distinct from the well known non-Abelian T-duality limit, is found.

\vfill

\end{titlepage}
\vfill
\eject



\def\baselinestretch{1.2}
\baselineskip 20 pt

\newcommand{\eqn}[1]{(\ref{#1})}

\tableofcontents

\section{Introduction}
\label{Sec:intro}

Two-dimensional coset conformal field theories (CFTs) \cite{coset} have traditionally played a significant r\^ole in the 
development of string theory. 
These models have proved suitable for describing consistent string propagation in curved spaces, since they have a Lagrangian realization in terms of gauged 
Wess--Zumino--Witten (WZW) models \cite{gwzw}. For this reason constructions based on non-compact 
groups have been of particular importance since they give rise to Minkowski signature curved metrics as well as to their Euclidean 
continuations.\footnote{For indicative references see \cite{wittenbh,Bars:1991pt,Fradkin:1991ie,Dijkgraaf:1991ba,Bars:1991zt, Horne:1991gn} and for even earlier works on the usefulness of non-compact current algebras in realizing curved space-time string models see \cite{Forgacs:1989ac,Petropoulos:1989fc,Bars:1989ph}. }

The relevant for the present work property of a class of gauged WZW models
based on non-compact groups is the fact that they have regions in their geometry which can be reached asymptotically. It turns out 
that these regions are described consistently by exact coset CFTs based on compact groups \cite{Petropoulos:2006py}.
The main conclusion of this work based on orthogonal groups is that at spatial infinity the geometry corresponding to the non-compact coset CFT  $SO(d,1)_{-k}/SO(d)_{-k}$ CFT is reduced to that of a compact coset CFT $SO(d)_k/SO(d-1)_k$ times a linear dilaton, 
where $k$ is the WZW level. We will take advantage of this specific feature in our construction 
of new integrable $\s$-models. In particular, recall the integrable $\s$-models constructed 
in \cite{Sfetsos:2013wia} via a gauging procedure resembling that of the gauged WZW models.
These models correspond to a finite deformation of CFT theories and form two basic classes. 
In the first class, in which most work in various directions has been done, belong deformations of WZW models by current bilinears.
In the second class they belong deformations of gauged WZW models by 
parafermion bilinears. Integrability of these models was shown for the group case in \cite{Sfetsos:2013wia} and for the coset one, when 
$G/H$ corresponds to a symmetric space in \cite{Hollowood:2014rla}.
We will focus on the second class of models for non-compact orthogonal 
groups such as the ones listed above.  
In the spirit of \cite{Petropoulos:2006py} we will consider $\l$-deformed models based on 
the non-compact coset CFT $SO(d,1)_{-k}/SO(d)_{-k}$. We will take the asymptotic limit which will
provide novel consistent integrable $\s$-models. They will interpolate between the compact coset CFT $SO(d)_k/SO(d-1)_k$ times a linear dilaton in the infrared (IR), and the hyperbolic space $H_d$ in the ultraviolet (UV). We will provide all details in the case of $d=2$ and $d=3$ for which the $\l$-deformed
backgrounds were constructed in full detail in \cite{Sfetsos:2013wia} and \cite{Demulder:2015lva}, respectively. In the simplest $d=2$ case it is possible to construct the model independently, without resorting to the asymptotic limit of the parent $\l$-deformed, but for $d=3$ and presumably for higher values of $d$, the limiting procedure seems imperative. We note that the running of the coupling under the renormalization 
group (RG) is unaffected by the asymptotic limit and is given precisely by the expression found in \cite{Itsios:2014lca,Sfetsos:2014jfa,Appadu:2015nfa}. 
We uncover the operator driving the perturbation away from the conformal point. In the parent theory this is in terms of the non-compact 
parafermion bilinears of  $SO(d,1)_{-k}/SO(d)_{-k}$ and is an irrelevant perturbation. However, for our models the interpretation should be
given instead in terms of fields of the compact coset theory $SO(d)_k/SO(d-1)_k$ which indeed we provide. 

Complementary to the class of examples described above, we will also consider examples with fixed points in the UV. 
Such models may be constructed by first recalling the $\l$-deformed coset CFTs $G_{k_1}\times G_{k_2}\ov G_{k_1+k_2}$ constructed in \cite{Sfetsos:2014cea} for $G=SU(2)$ and in \cite{Sfetsos:2017sep} for general compact groups.
These  are integrable for arbitrary values of the levels $k_1$ and $k_2$. This construction may 
be easily adopted to non-compact groups in which asymptotic limits can be taken. We provide an explicit example 
for $G=SL(2,\IR)$ which 
gives rise to a smooth interpolation between two spacetimes with cosmological and black hole interpretations.

The rest of the paper is as follows: We begin in Section \ref{Sec:2} by studying the two-dimensional integrable model interpolating between
the Euclidean space $E_2$ with a linear dilaton in the IR and the hyperbolic space $H_2$ in the UV. In Section \ref{Sec:3}, we study the three-dimensional integrable model  interpolating between the $SO(3)_k/SO(2)$ times a linear dilaton in the IR and the hyperbolic space $H_3$ in the UV. In Section \ref{Sec:4}, we construct a three-dimensional integrable model which interpolates between two exact coset CFTs times a linear dilaton. We provide concluding remarks in Section \ref{Sec:5}.

\section{The two-dimensional integrable model}
\label{Sec:2}

In this section we present the simplest example with a $\s$-model having a two-dimensional target space and we show that it interpolates between
the Euclidean space $E_2$ with a linear dilaton in the IR and the hyperbolic space $H_2$ in the UV.

Consider the $\s$-model action for two fields $\r$ and $\varphi$ given by\footnote{The worldsheet $(\tau,\sigma)$ and light-cone coordinates $\s^\pm$ are given by
\begin{equation*}
\s^\pm=\tau\pm\s\,,\quad\del_\tau=\del_++\del_-\,,\quad\del_\s=\del_+-\del_-\,,\quad \text{d}^2\s=\text{d}\tau\,\text{d}\s\,.
\end{equation*}}
\begin{equation}
\label{2daction}
\begin{split}
S=  {k\ov \pi}  \int \text{d}^2\s\, &  \Bigg[\bigg(\frac{1 +\l^2}{ 1 - \l^2}  +  \frac{2\l}{ 1 - \l^2} \cos 2\varphi\bigg) \del_+\r\del_-\r
 \\
 & + \bigg(\frac{1 +\l^2}{ 1 - \l^2} -  \frac{2\l}{ 1 - \l^2} \cos 2\varphi\bigg) \del_+\varphi\del_-\varphi
 \\
 &
\quad  - \frac{2  \l}{1 - \l^2} \sin2\varphi \, (\del_+\r \del_-\varphi + \del_+\varphi \del_-\r) \bigg) \Bigg]\, ,
\end{split}
\end{equation}
where $\l\in[0,1)$ and $k\in \mathbb{Z}^+$ are parameters. These ranges warrant Euclidean signature. 
In addition, this $\s$-model is non-singular.

\no
Due to the isometry corresponding to shifts of $\r$ one may also consider the $T$-dual to the $\s$-model \eqn{2daction}.
 However, it turns out that, using an appropriate coordinate transformation, this action is self-dual under T-duality. 
Moreover, at the end of the following subsection we will explain what singles 
out  the form of the above action having just one coupling constant parametrized by $\l$.

\subsection{Renormalizability}

We first investigate the renormalizability of the model by CFT methods. 
We begin by introducing new fields $\phi_1$ and $\phi_2$ via the rescaling
\be
\label{eq.change}
\r= {\phi_1\ov \sqrt{2 k}}  \ ,\qq \varphi= {\phi_2\ov \sqrt{2 k}} \ .
\ee
Then expanding the action for small $\l$, we have that $S=S_0 + S_1 +\dots $. Moving to the Euclidean regime,\footnote{In passing to the Euclidean regime we use the analytic continuation
\begin{equation*}
\begin{split}
& \tau= -i t\,,\quad
\s^+ = - i z\ ,\quad \s^- = - i \z\ ,\quad   \del_+= i \del\ ,\quad \del_-= i \bar \del\,,\quad \text{d}^2z=\text{d}t\,\text{d}\s\,,
\end{split}
\end{equation*}
where $z= t+i \s$ and  $\z= t-i \s$. In addition, the action in the Euclidean path integral appears as $\text{e}^{-S}.$
\label{Analytic.continue}
} 
the leading term is given by
\be
\label{leda}
S_0= {1\ov 2\pi} \int \text{d}^2z\, (\del\phi_1 \bar\del \phi_1 + \del\phi_2 \bar\del \phi_2  )\ ,
\ee
that is two-dimensional flat space. These fields have propagators given by\footnote{The propagator is read as usual by varying \eqref{leda},
with the propagator $G_{ij}=\langle \phi_i(z,\z)\phi_j(w,\bar w) \rangle$ obeying $\del\bar\del G_{ij}=-\pi\d_{ij} \d^{(2)}(z-w)$, with the $\d$-function arising from $\del\frac{1}{\z}=\bar\del\frac{1}{z}=\pi\d^{(2)}(z)$, where $\d^{(2)}(z)=\d(t)\d(\s)$.\label{norm.propagator}}
\be
\langle \phi_i(z,\z)\phi_j(w,\bar w) \rangle= -\d_{ij} \ln |z-w|^2\,,\quad i=1,2\, .
\ee
The first correction to $S_0$ is
\be
\begin{split}
S_1 = {\l \ov \pi}  \int \text{d}^2z\,  & \bigg[\cos\Big(\sqrt{2\ov k} \phi_2\Big)\,  (\del\phi_1\bar\del\phi_1 -  \del\phi_2\bar\del\phi_2)
\\
& -   \sin\Big(\sqrt{2\ov k} \phi_2\Big)\, (\del\phi_1\bar\del\phi_2 + \del\phi_2\bar\del\phi_1) \bigg]\ ,
\end{split}
\ee
which consists of operators of the type $\del\phi_i\bar\del\phi_j$, dressed with the exponential factors $\text{e}^{\pm i \sqrt{\nicefrac2k}\,\phi_2}$,\,  
of conformal dimension equal to $h=\bar h = \nicefrac{1}{k}$. 
Hence, all terms of the perturbation in $S_1$ have scaling dimension
\be
\D = 2+h+\bar h =2 +{2\ov k}\ .
\ee
Therefore, the $\beta$-function to leading order in $\l$ is given by
\be
\label{beta2}
{\text{d}\l \ov \text{d}t} = {\D-2\ov 2} \l = {\l\ov k}\, ,
\ee
where $t=\ln \m^2$, $\mu$ is the energy scale.
One may expect higher order corrections in $k$ and $\l$. We will show that to leading order in $k$, this result is exact in $\l$.  This
will be done using the RG flow equations for the $\s$-model action
\be
\label{norm.action}
S={1\ov 2\pi} \int \text{d}^2\s\, g_{\m\n}\del_+ X^\m\del_- X^\n\,,
\ee
whose $\b$-functions to lowest order in the curvature read \cite{Ecker,Friedan:1980jf}
\begin{equation}
 \label{RGflow}
 \frac{\text{d} g_{\m\n}}{\text{d}t} = R_{\m\n} + \nabla_\mu\xi_\nu+ \nabla_\nu\xi_\mu\ ,
\end{equation}
where $\xi$'s correspond to diffeomorphisms.
For the $\s$-model \eqn{2daction} this implies exactly \eqn{beta2} provided that
\begin{equation}
\label{diff}
 \xi_\m = \del_\m \Phi \, ,
\end{equation}
with
\be
\label{dil.sec2}
\Phi=-\r\ .
\ee
Curvature corrections are small provided that $k\gg 1$.

\no
It is worth noting that, if we had started with arbitrary coupling coefficients in the various terms in \eqn{2daction}, renormalizability
would have fixed them to the given expressions. 
The $\b$-function \eqn{beta2} implies the IR fixed point at $\l=0$ at which the $\s$-model is a free theory with a linear dilaton.
In addition, towards the UV, at some scale, one reaches $\l=1$. This scale can be pushed to arbitrary large values provided that $k$ is sent to infinity.
We are familiar with a similar procedure from the so-called non-Abelian limit of $\l$-deformed models \cite{Sfetsos:2013wia}. 
We will see that the space in this deep UV limit is nothing but $H_2$, but the associated zoom-in limit is different. 

\subsection{ Interpolating between $E_2$ and $H_2$}
\label{Interpolating.2d}

The action \eqn{2daction} is invariant under the following non-perturbative symmetries in the space of couplings $k$ and $\l$
\be
\label{III}
({\rm I}):\quad  \l\to {1\ov \l} \ ,\qq k\to -k\ , \qq ({\rm II}):\quad  \l\to -\ \l \ ,
\ee
where the latter should be accompanied with the coordinate shift $\varphi \to \nicefrac\pi2 + \varphi$. Note that both symmetries leave
the $\b$-function \eqn{beta2} invariant.

In what follows we will show that the model \eqn{2daction} interpolates between the
$\s$-models corresponding to two-dimensional  Euclidean space $E_2$ including a linear dilaton in the IR and the hyperbolic space $H_2$ in the UV. 
In the IR, $\l=0$ and the action \eqref{2daction} drastically simplifies to the two-dimensional  Euclidean space $E_2$ 
\be
\label{2daction.IR}
S_\text{IR}=\frac{k}{\pi}\int\text{d}^2\s\left(\del_+\rho\del_-\rho+\del_+\varphi\del_-\varphi\right)\,,
\ee
and the linear dilaton is given by \eqref{dil.sec2}. In the UV, we perform the following zoom-in
\be
\label{dif.lambda.one}
\l=1-\frac{\k^2}{k}\,,\quad \varphi=\frac{\pi}{2}-\frac{\k^2}{2k}x\,\text{e}^{-\rho}\,,\quad k\gg1\,,
\ee
yielding the hyperbolic $H_2$ space
\be
\label{H2.examp2d}
S_\text{UV}=\frac{\k^2}{2\pi}\int\text{d}^2\s\,\left(\del_+\rho\del_-\rho+\text{e}^{-2\rho}\del_+x\del_-x\right)
\ee
and the dilaton is given by \eqref{dil.sec2}.

\subsubsection*{Central charge of the IR CFTs \eqn{2daction.IR} and \eqn{dil.sec2}}

Note that the flat space CFT has a linear dilaton background, 
corresponding to a background charge $Q$, i.e. $\Phi= Q X$. 
Since it is important to set up the normalizations correctly we will provide some details by
first recalling that for the linear dilaton background the Euclidean action (see footnote~\ref{Analytic.continue}) in a worldsheet with metric $\g_{ab}$ is given by\footnote{In a generic background, the dilaton is specified by the following one-loop equation~\cite{Callan:1985ia}
\begin{equation*}
S_{\ell.\text{d}}= {1\ov 2 \pi \a'} \int \text{d}^2z\, \sqrt{\g} \big(\g^{ab}(g_{\mu\nu}+B_{\mu\nu})\del_a X^\mu \del_b X^\nu + \a' \Phi R_\g\big)\,,\quad
R^-_{\mu\nu}+2\nabla^-_\mu\del_\nu\Phi=0\,,
\end{equation*}
where $g_{\mu\nu}$ and $B_{\mu\nu}$ is the metric and the two-form.
The torsionfull covariant derivative, Riemann and Ricci tensors are constructed out of the connection 
$\Gamma^-_{\mu\nu}{}^\rho=\Gamma_{\mu\nu}{}^\rho-\frac12 H_{\mu\nu}{}^\rho$, where $H_{\mu\nu\rho}$ is the field strength of $B_{\mu\nu}$, that is
$$\nabla^-_\mu V_\nu=\del_\mu V_\nu-\G^-_{\mu\nu}{}^\rho V_\rho\,,\quad 
[\nabla^-_\mu,\nabla^-_\nu]V_\rho=R^-_{\mu\nu\rho}{}^\s V_\s-H_{\mu\nu}{}^\s\nabla^-_\s V_\rho\,,\quad R^-_{\mu\nu}=R^-_{\mu\rho\nu}{}^\rho\,.$$
}
\be  
\label{action.linear.dilaton}
S_{\ell.\text{d}}= {1\ov 2 \pi \a'} \int \text{d}^2z\, \sqrt{\g} \big(\g^{ab}\del_a X \del_b X + \a' \Phi R_\g\big)\,,
\ee
where $R_\g$ is the Ricci scalar built out of the worldsheet metric $\g_{ab}$.
The second term in \eqref{action.linear.dilaton} vanishes in a flat worldsheet (where $\g_{z\z}=\d_{z\z}=\nicefrac12$)
\be
\label{Sld.flat}
S_{\ell.\text{d}}={1\ov \pi\a'}  \int \text{d}^2z\, \del X \bar \del X\, ,
\ee
but it alters the stress--energy tensor which is defined by $\displaystyle T_{ab}= -{2\pi\ov\sqrt{\g}} {\d S_{\ell.\text{d}}\ov \d \g^{ab}}\bigg{|}_{\g_{ab}=\d_{ab}}$, so that
\be 
T_{zz}= -{1\ov \a'} (\del X)^2 + Q\, \del^2 X\,, \quad  T_{z\z}=-Q\del\bar\del X\,.
\ee
Note that the stress--energy tensor is trace-full, the $T_{z\z}$ component is non-vanishing, as expected since the dilaton coupling in \eqn{action.linear.dilaton} is not Weyl invariant.
Using the two-point function $ \langle X(z) X(w)\rangle = -\nicefrac{\a'}{2} \ln|z-w|^2$, obtained by \eqref{action.linear.dilaton}, the central charge is given by
\be
\label{gen.central.lin}
c_{\ell.\text{d}}= 1+ 6 \a' Q^2\  .
\ee  
Therefore comparing with \eqref{dil.sec2} and \eqref{2daction.IR} we 
identify $\rho$ with $X$, yielding $\a'=\nicefrac1k$ and background charge  $Q= -1$. Then, the central charge \eqref{gen.central.lin} is 
\be
\label{central.2d.dilaton}
c_{\ell.\text{d}} = 1 +{6\ov k}\ .
 \ee
The original theory at the CFT point is $SL(2,\IR)_{-k}/SO(1,1)$ and has central charge
\be
c_{\rm coset}= {3k\ov k-2}-1= 2 +{6\ov k} + \dots\ .
\ee
Therefore,  $c_{\rm coset}= 1 +  c_{\ell.\text{d}}$, as it should be.

\no
In summary, we have proved that the model \eqref{2daction} interpolates under the renormalization group flow between two target spaces, namely a flat CFT \eqref{2daction.IR} with a linear dilaton background \eqref{dil.sec2} and the hyperbolic $H_2$ space \eqref{H2.examp2d} in the IR and towards the UV, respectively.

\subsection{Integrability}
\label{Sec:Integrability}

In order to prove integrability we will write the equations of motion in the Lax formalism. The equations of motion arising from the variations 
of $\varphi$ and $\r$ in the action \eqref{2daction} are, respectively,
\begin{equation}
 \label{eqsofmotiontr}
 \begin{aligned}
  & 2 \l\, \sin2\varphi \, \big( \partial_+ \r \, \partial_- \r + \partial_+ \varphi \, \partial_- \varphi - \partial_+ \partial_- \r \big) + W_- \, \partial_+ \partial_- \varphi = 0 \, ,
  \\[5pt]
  & \partial_+ \big( W_+ \, \partial_- \r - 2 \l \, \sin 2 \varphi \, \partial_- \varphi \big) + \partial_- \big( W_+ \, \partial_+ \r - 2 \l \, \sin 2 \varphi \, \partial_+ \varphi \big) = 0 \, ,
 \end{aligned}
\end{equation}
where we have set
\begin{equation}
 W_\pm = 1 + \l^2 \pm 2 \, \l \, \cos 2 \varphi \, .
\end{equation}
Next we define
\begin{equation}
\label{LAX.asymp}
 \begin{aligned}
  & \cL^1_\pm = \zeta^{\pm 1} \frac{\sqrt{\l}}{1 + \l} \big( \sin \varphi \, \partial_\pm \r + \cos \varphi \, \partial_\pm \varphi \big)  \, ,
  \\
  & \cL^2_\pm = \pm \zeta^{\pm 1} \frac{\sqrt{ \l}}{1 - \l} \big( \cos \varphi \, \partial_\pm \r - \sin \varphi \, \partial_\pm \varphi \big)\, ,
  \\
  & \cL^3_\pm = \pm \frac{i}{2 (1 - \l^2)} \big( 2 \, \l \, \sin 2 \varphi \, \partial_\pm \r
  - W_- \, \partial_\pm \varphi \big)  \, ,
 \end{aligned}
\end{equation}
where $\zeta$ is a complex parameter, as well as the matrix-valued one-form
\begin{equation}
 \cL = \cL_+ \, \text{d}\s^+ + \cL_- \, \text{d}\s^- \, , \qquad \cL_\pm = \sum_{a=1}^3\cL^a_\pm \, \s_a \,,
\end{equation}
where $\s_a $ are the Pauli matrices.
Then $\cL$ satisfies the flatness condition
\begin{equation}
 \text{d} \cL = \cL \wedge \cL  \, ,
\end{equation}
provided that the equations of motion \eqn{eqsofmotiontr} are satisfied. Hence, the above $\cL_\pm$ constitute a Lax pair in which
$\zeta$ plays the r\^ole of the spectral parameter, proving the integrability of the model~\eqref{2daction} in the weak sense. 
To prove the integrability in the Hamiltonian sense we need to show that the conserved charges are in involution. More precisely, we need to show that the 
spatial Lax component, namely ${\cal L}_\s={\cal L}_+-{\cal L}_-$ of \eqref{LAX.asymp}, obeys the Maillet $r/s$-form~\cite{Sklyanin:1980ij,Maillet:1985ek,Maillet:1985ec}.
However, it turns out that this property is inherited from the $\l$-deformed $SL(2,\IR)_{-k}/SO(1,1)$ $\s$-model, whose integrability in the weak and in the Hamiltonian sense were demonstrated in~\cite{Hollowood:2014rla} and~\cite{Hollowood:2015dpa} 
respectively, and its connection with the model~\eqref{2daction} will be established in the next subsection.

\subsection{Connection to the $\l$-deformed models}

In what follows we will connect the model~\eqref{2daction} with the $\l$-deformed model based on the $SL(2,\IR)_{-k}/SO(1,1)$ exact CFT
\begin{equation}
\label{ch2}
\begin{split}
 S= & {k\ov \pi}  \int \text{d}^2\s\,  \bigg( \frac{1 - \l}{ 1 + \l} \big(   \del_+\r\del_-\r  + \coth^2 \r \, \del_+\varphi\del_-\varphi \big)
 \\
 &
 + \frac{4  \l}{1 - \l^2} \big(  \cos\varphi \, \del_+\r - \sin\varphi \, \coth\r \, \del_+\varphi  \big)\big(\cos\varphi \, \del_-\r - \sin\varphi \, \coth\r \, \del_-\varphi  \big)\bigg) \,.
\end{split}
\end{equation}
This is the non-compact analogue of the original $\l$-deformed model  based on the $SU(2)_k/U(1)$ coset CFT \cite{Sfetsos:2013wia}.
These two models, namely \eqref{2daction} and \eqref{ch2}, have various properties in common.
Firstly, they obey the same non-perturbative symmetries \eqref{III}, which for the action  \eqref{ch2} were found in~\cite{Itsios:2014lca,Georgiou:2020bpx}. 
Secondly, they share the same $\b$-function \eqref{beta2}, which for the action \eqref{ch2}  was found  in \cite{Itsios:2014lca}, up to a
flip of sign for $k$ and the  diffeomorphism is given by \eqn{diff} with
\begin{equation}
\label{DilatonAdS2La}
  \Phi = - \ln \sinh\rho \, .
\end{equation}
This is the dilaton arising from integrating out non-dynamical gauge fields in the construction of the model. 
Thirdly, these models are classically integrable and their equations of motion take the form of the Lax pair. 
The integrability of the model~\eqref{2daction} was demonstrated in subsection~\ref{Sec:Integrability}
and of the model \eqref{ch2} its weak and Hamiltonian integrability were shown in~\cite{Hollowood:2014rla} and~\cite{Hollowood:2015dpa}, respectively.

\no
Finally, there exists an analogue zoom-in limit \eqref{dif.lambda.one} in the $\l$-deformed $SL(2,\IR)_{-k}/SO(1,1)$ action
\eqref{ch2}, \eqref{DilatonAdS2La}. In particular, we consider
\be
\label{dif.lambda.two}
\l=1-\frac{\k^2}{k}\,,\quad \varphi=\frac{\pi}{2}-\frac{\k^2}{4k}\frac{x}{\cosh\rho}\,,\quad k\gg1\,,
\ee
yielding  the Abelian T-dual of the hyperbolic $H_2$
\be
S_\text{UV}=\frac{\k^2}{2\pi}\int\text{d}^2\s\left(\del_+\rho\del_-\rho+\frac{1}{4\sinh^2\rho}\del_+x\del_-x\right)\,,
\ee
whereas the dilaton is given by \eqref{DilatonAdS2La}. 
Note that the zoom-in limit \eqref{dif.lambda.two} we have taken is a different limit compared to the usual non-Abelian T-dual of the
hyperbolic $H_2$ space around $\l=1$. In the latter limit, one zooms-in around the identity group element  ($\rho=\varphi=0$)~\cite{Sfetsos:2013wia}.

\no
The above properties in common are not a coincidence as the action \eqn{2daction} and the scalar \eqref{dil.sec2} are obtained asymptotically for $\r\gg 1$, from 
the action \eqref{ch2} and the scalar \eqref{DilatonAdS2La} respectively. Finally, the Lax connections match in the asymptotic limit and the action \eqref{2daction} is also integrable in the Hamiltonian sense.

\section{ Higher dimensional models }
\label{Sec:3}

We are interested in higher dimensional analogues of the previous example. Clearly it is rather cumbersome 
to start with an educated ansatz similar to that in \eqn{2daction}. In this section we proceed with the following observation. 

\no
We recall that the asymptotic behaviour of the $\l$-deformed model \eqn{ch2} is well defined and additionally that
this has already an analogue in a class of $\s$-models corresponding to exact CFTs. 
Specifically, in \cite{Petropoulos:2006py} it was shown that the background fields corresponding to the
exact CFT coset model $SO(d,1)_{-k}/SO(d)_{-k}$ asymptotically become the ones for the $SO(d)_k/SO(d-1)_k$ exact CFT times
a linear dilaton. This correspondence is delicate and takes into account an appropriate shift of the level $k$ and
a precise value for the background charge corresponding to the linear dilaton. For the reader's convenience we provide the relation 
for the central charges
\be 
\begin{split}
 SO(d,1)_{-k}/SO(d)_{-k}&:\qq c_{d,k}= {d(d+1) k\ov 2k - (d-1)}- {d(d-1) k\ov 2k - (d-2)}\ ,
\\
  SO(d+1)_{k}/SO(d)_{k}& :\qq  \tilde c_{d,k}= {d(d+1) k\ov 2k + (d-1)}- {d(d-1) k\ov 2k + (d-2)}\ . 
\end{split}
\ee 
Clearly
\be 
\label{lin.dilaton.gen}
c_{d,k+d-2} - \tilde c_{d-1,k} = 1+ {3(d-1)^2\ov 2k +d-3}\ ,  
\ee
is the central charge corresponding to the linear dilaton.\footnote{The expressions \eqref{central.2d.dilaton}, \eqref{lin.dilaton.gen}
match for $d=2$ and large level $k$. This agreement relies on the fact that the quadratic Casimir operators in adjoint representation of the $SL(2,\IR)$ and the $SO(3)$ are four and one respectively. Effectively this is equivalent in sending the level $k$ to level $4k$ in \eqref{central.2d.dilaton}.} 
For details as well as similar asymptotic limits for other coset CFTs we refer the reader to \cite{Petropoulos:2006py}. 
Given the above, one may start with the $\l$-deformed model
based on the coset CFT $SO(d,1)_{-k}/SO(d)_{-k}$ and then take the asymptotic limit. 
Though not a priori obvious, it turns out that this is well defined.
We will present all details for the $d=3$ case.

\subsection{The three-dimensional integrable model}

The background fields corresponding to the non-compact coset 
$SO(3,1)_{-k}/SO(3)_{-k}$ CFT were constructed in  \cite{Bars:1991pt, Fradkin:1991ie}. They comprise  
the action \eqref{norm.action}  whose target space is described by the metric
\be
\label{met3d}
\text{d}s^2  = 2 k \bigg( \frac{\text{d}b^2}{b^2-1} + \frac{1 + b}{1 - b} \frac{\text{d}v^2}{v (v - u - 2)}
 - \frac{1 - b}{1 + b}  \frac{\text{d}u^2}{u (v - u - 2)}\bigg)\, .
\ee
There is also the dilaton
\be
\label{3ddil}
\text{e}^{-2 \Phi}  = (b^2 - 1) (v - u - 2)\ .
\ee
The range of variables giving rise to a Euclidean signature background is
\be
\label{rangee}
|b| > 1\ ,\qq 0 < v < u + 2 < 2\ .
\ee
The above global coordinate system, found in \cite{Bars:1992ti}, has the advantage that by changing the ranges and/or flipping the sign of $k$ accommodates all backgrounds corresponding to compact and non-compact coset CFTs with Euclidean and Minkowski signature, related to $SO(3,1)_{-k}/SO(3)_{-k}$.

\no
The $\l$-deformed model based on the above CFT was constructed in \cite{Demulder:2015lva}. In the above coordinates there is a metric
\ba
\label{threml}
&&\text{d}s^2  =   2k  \bigg( \frac{1 + \lambda^2 + 2 \lambda (v - u - 1)}{1 - \lambda^2} \frac{\text{d}b^2}{b^2-1} 
+ \frac{1 + b}{1 - b} \frac{1 + \lambda^2 + 2 \lambda (1 - v)}{1 - \lambda^2} \frac{\text{d}v^2}{v (v - u - 2)}
\nonumber
\\
&&\qquad - \frac{1 - b}{1 + b} \frac{1 + \lambda^2 + 2 \lambda (1 + u)}{1 - \lambda^2} \frac{\text{d}u^2}{u (v - u - 2)} - 
\frac{4 \lambda}{1 - \lambda^2} \frac{\text{d}u \, \text{d}v}{v - u - 2}
\\
&&\qquad  - \frac{4 \lambda}{1 - \lambda^2} \frac{\text{d}b \,\text{d}u}{1 + b} - \frac{4 \lambda}{1 - \lambda^2} \frac{\text{d}b \,\text{d}v}{1 - b} \bigg)
\nonumber
\ea
and the same dilaton as that in \eqn{3ddil}. Clearly, for $\l=0$ one obtains \eqn{met3d}. The perturbation driving the model away from the
conformal point is bilinear in terms of parafermionic operators of the $SO(3,1)_{-k}/SO(3)_{-k}$ coset CFT. The perturbation is irrelevant 
due to the fact that these are non-compact and have conformal dimension in excess of unity.\footnote{Quantum properties 
of non-compact parafermions have been studied thoroughly only for those corresponding to the lower dimensional coset $SL(2,\IR)_{-k}/U(1)$ \cite{Lykken:1988ut}. These are clearly distinct from their compact counterparts for the coset $SU(2)_k/U(1)$ of \cite{Fateev:1985mm}.}
This is encoded in the $\beta$-function we turn to below.

\no
The RG flow equation \eqn{RGflow} implies that
\begin{equation}
 \label{betaFunclambda}
 \frac{\text{d}\l}{\text{d}t} = \frac{\l}{2 k} \, ,
\end{equation}
provided that the diffeomorphism is given by \eqn{diff}.  
Note that both symmetries in \eqn{III} are respected by the $\b$-function
\eqn{betaFunclambda}. 
However, only (I) is a manifest symmetry of the $\l$-deformed metric.
The reason  why the symmetry (II) is not manifest will be explained below.

\subsection{Taking the asymptotic limit}

In order to take the asymptotic limit in the spirit of \cite{Petropoulos:2006py} we first perform the coordinate transformation
\begin{equation}
\label{buv}
\begin{split}
& u = -2 \sin^2\th \, \cos^2\phi \, , \qquad v =2 \sin^2\th \, \sin^2\phi\ ,
\\
&
b = \frac{(1 - \l)^2 \text{e}^{2 \rho} + 4 \, \l \, \cos 2 \phi \, \sin^2\th}{W} \, , \qquad  W= 1 + \l^2 - 2 \, \l \, \cos 2\th \, ,
\end{split}
\end{equation}
which is in agreement with the ranges of variables in \eqn{rangee}, including the variable $b$ for large values of $\r$ for which we are interested in.
The resulting geometry has the following limit for large $\rho$
\begin{equation}
 \label{decoupledMetric}
 \text{d}s^2 = 8 k \bigg( \frac{W}{1 - \l^2} \text{d}\rho^2 + \frac{1 - \l^2}{W} \text{d}\th^2 + \frac{1 + \l}{1 - \l} \tan^2\th \, \text{d}\phi^2 \bigg) \, ,
\end{equation}
whereas the dilaton becomes
\begin{equation}
 \label{decoupledDila}
 \Phi = - 2 \rho - \ln \frac{\cos\th}{W} \, ,
\end{equation}
where we have omitted a constant additive term and accordingly in analogue expressions below.

\no
The RG flow equation \eqref{RGflow} implies again \eqref{betaFunclambda} where now the diffeomorphism reads
\begin{equation}
 \xi_\m =\del_\m (\Phi+2\rho) \ ,
\end{equation}
that is only the $\th$-dependent part of the dilaton \eqn{decoupledDila} plays a r\^ole in the 
diffeomorphism.

\no
Note that the symmetry (I) in \eqn{III} is manifest in the metric \eqn{decoupledMetric} and the dilaton \eqn{decoupledDila}.
It can be easily seen that the symmetry (II) accompanied by the replacement $\th\to \nicefrac{\pi}{2}-\th$ gives the metric
 \begin{equation}
 \label{decoupledMetricv}
 \text{d}s^2 = 8 k \bigg( \frac{W}{1 - \l^2} \text{d}\rho^2 + \frac{1 - \l^2}{W} \text{d}\th^2 + \frac{1 - \l}{1 + \l} \cot^2\th \, \text{d}\phi^2 \bigg) \, 
\end{equation}
and the dilaton
\begin{equation}
 \label{decoupledDilav}
 \Phi = - 2 \rho - \ln \frac{\sin\th}{W} \, .
\end{equation}
 These correspond to the T-dual model of \eqn{decoupledMetric} and \eqn{decoupledDila},
 with respect to the cyclic coordinate $\phi$.
Hence, (II) of \eqn{III} is a symmetry in the space of T-dual equivalent models. Since \eqn{decoupledMetric} originates from \eqn{threml}, this is in accordance with the fact that the symmetry (II) is not leaving \eqn{threml} invariant. In that respect note that the latter metric does not have isometries. Let us note that the $\beta$-function has more symmetries than the action it is derived from. This property was also noted in a subclass of the integrable models constructed in \cite{Georgiou:2017aei,Georgiou:2017jfi}, where (I) of \eqref{III} is manifest at the level of the beta function but not in the action.

\no
The weak and the Hamiltonian integrability as well as the renormalizability of our model stem from that of the $\l$-deformed $SO(3,1)_{-k}/SO(3)_{-k}$ model, the latter being based on a symmetric space \cite{Hollowood:2014rla} and \cite{Hollowood:2015dpa} as well as \cite{Itsios:2014lca,Sfetsos:2014jfa,Appadu:2015nfa} respectively.

\subsubsection{From $SO(3)_k/SO(2)$ times a linear dilaton towards $H_3$}

In the IR, for $\l=0$ the metric \eqn{decoupledMetric} and the dilaton \eqn{decoupledDila} correspond to the exact coset CFT $SO(3)_{k}/SO(2)$ times
a linear dilaton background
\be
\label{3d.IR}
S_\text{IR}=\frac{4k}{\pi}\int\text{d}^2\s\left(\del_+\rho\del_-\rho+\del_+\th\del_-\th+\tan^2\th\del_+\phi\del_-\phi\right)\,.
\ee
The dilaton 
\be
\label{3d.IR.dil}
 \Phi = - 2 \rho - \ln\cos\th,
\ee
is also in accordance with the results of  \cite{Petropoulos:2006py}.
The central charge at the conformal point can be obtained by adding the two-contributions,
namely that of the exact coset CFT and of the linear dilaton background. The central charge of the coset CFT is given by
\be
c_\text{coset}=\frac{6k}{2k+1}-1=2-\frac{3}{2k}+{\cal O}\left(\frac{1}{k^2}\right).
\ee
To evaluate the dilaton's central charge \eqref{gen.central.lin} we follow
the discussion at the end of Section~\ref{Interpolating.2d}, 
identifying $X$ with $\rho$, the background charge $Q=-2$ and $\a'=\frac{1}{4k}$. This gives
\be
c_{\ell.\text{d}}=1+\frac{6}{k}\, ,
\ee
matching the general result \eqref{lin.dilaton.gen} for $d=3$ and for large level $k$.
Employing the above we find
\be
c_\text{IR}=c_\text{coset}+c_{\ell.\text{d}}=3+\frac{9}{2k}+{\cal O}\left(\frac{1}{k^2}\right)\, ,
\ee
which is the correct expression for the non-compact coset $SO(3,1)_{-k}/SO(3)_{-k}$ CFT in the large $k$ limit.

In the UV, we perform the following zoom-in limit
\be
\l=1-\frac{\k^2}{k}\,,\quad \th=\frac{\k^2}{2k}\sinh\a\,,\quad k\gg1
\ee
and we find the hyperbolic $H_3$ space
\be
\label{H3.examp3d}
S_\text{UV}=\frac{2\k^2}{\pi}\int\text{d}^2\s\left(\del_+\a\del_-\a+\cosh^2\a\del_+\rho\del_-\rho+\sinh^2\a\del_+\phi\del_-\phi\right)\,,
\ee
whereas the dilaton equals to
\be
\Phi=-2\rho+2\ln\cosh\a\,.
\ee
Hence, similarly to the $d=2$ case we get in the above zoom-limit a hyperbolic
space, instead of the non-Abelian T-dual of the principal chiral model for the coset $SO(3,1)/SO(3)$.

\no
In brief, we have demonstrated that the model \eqref{decoupledMetric}, obtained from \eqref{threml} via the asymptotic limit \eqref{buv}, interpolates under the renormalization group flow between two target spaces, namely an exact coset CFT $SO(3)_{k}/SO(2)$ background \eqref{3d.IR} times a linear dilaton \eqref{3d.IR.dil} and the hyperbolic $H_3$ space \eqref{H3.examp3d} in the IR and towards the UV, respectively.

\subsubsection{Operator content of the perturbation}

The linear in $\l$ correction of \eqref{3d.IR}  is 
\be
\label{s1c}
S_1 = {8k\l\ov \pi} \int \text{d}^2\s\, \big( \cos 2 \th (\del_+\th \del_-\th - \del_+\rho\del_-\rho) + \tan^2\th \, \del_+\phi\del_-\phi \big)  \ .
\ee
We would like to interpret this in terms of a perturbation by operators of the underlying coset CFT. The perturbation should be organized using parafermion bilinears as well as primary operators of the same CFT. 

\no
Consider an element $g$ of the $SU(2)\simeq SO(3)$ group parametrized as
\begin{equation}
 g = \text{e}^{i \frac{\phi + \psi}{2} \s_3}  \text{e}^{i (\frac{\pi}{2} - \th)\s_2}  \text{e}^{i \frac{\phi - \psi}{2} \s_3}=  
 \begin{pmatrix}\text{e}^{i\phi}\sin\th & \text{e}^{i\psi}\cos\th\\ -\text{e}^{-i\psi}\cos\th & \text{e}^{-i\phi}\sin\th\end{pmatrix} \, .
\end{equation}
The adjoint action with respect to the group element $g$ is realized through the orthogonal matrix $D_{ab}$ defined as
\begin{equation}
 D_{ab} = \ha \tr(\s_a g \s_b g^{- 1}) \,.
\end{equation}
The classical parafermions are defined as \cite{Bardacki:1990wj}
\begin{equation}
\begin{split}
  & \psi_\pm = \pm \text{e}^{\mp i (\phi + \psi)} \big( \del_+ \th \mp i \tan \th \, \del_+ \phi \big) \, ,
  \\
  & \bar{\psi}_\pm = \pm \text{e}^{\pm i (\phi - \psi)} \big( \del_- \th \pm i \tan \th \, \del_- \phi \big) \, 
 \end{split}
\end{equation}
and are the classical counterparts of the compact CFT parafermions of \cite{Fateev:1985mm}. The variable
$\psi$ drops out in the $\s$-model due to the gauge invariance. However,  it appears
as a non-local phase  in the parafermions and depends on the fields $\th$ and $\phi$. 
This phase is necessary to make them chiral and anti-chiral \cite{Bardacki:1990wj}. 
Its precise form is not important, but it should drop out when the parafermions are combined in 
bilinears, possibly also dressed with additional operators. Then, it is easy to show that the correction \eqn{s1c} becomes
\be
\label{s1c1}
S_1 = {8k\l\ov \pi} \int \text{d}^2\s\, \bigg[ - \cos 2 \th\,   \del_+\rho\del_-\rho +  \big(\psi_+ \  \psi_-\big) M
  \begin{pmatrix}
   \bar{\psi}_+
   \\
   \bar{\psi}_-
  \end{pmatrix} \bigg]\ ,
\ee
where the matrix 
\begin{equation}
 M = \frac{1}{2}
 \begin{pmatrix}
  \text{e}^{2 i \psi} \cos^2 \th & \text{e}^{2 i \phi} \sin^2 \th
  \\
  \text{e}^{- 2 i \phi} \sin^2 \th & \text{e}^{- 2 i \psi} \cos^2 \th
 \end{pmatrix} \, .
 \ee
We consider primary fields of the coset CFT with equal holomorphic and antiholomorphic dimensions given by
\be
h_{j,m}= \bar h_{j,m}= {j(j+1)\ov 4 k + 2}- {m^2\ov 4k}\ ,
\ee
where $j$ labels the $SO(3)$ representation to which the operator belongs to and $m$ is its charge.
The matrix elements $D_{ab}$ belong to the representation with $j=1$, but only linear combinations of them have specific charges, 
with values $m=-1,0,1$. 
Explicitly $\cos2\th = - D_{33}$, has $m=0$. Then $\displaystyle h_{1,0}= \frac{1}{2k}$, for large $k$ and the operator 
corresponding to the first term in \eqn{s1c1} has scaling dimension $\displaystyle 2(1+h_{1,0})= 2+\nicefrac1k$. The same of course 
should be true for the second term  in \eqn{s1c1} as well, but this requires a further explanation.
The matrix $M$ can be written in terms of the elements of the matrix $D_{ab}$ as
\begin{equation}
 M = \frac{1}{4}
 \begin{pmatrix}
  D_{22} - D_{11} + i (D_{12} + D_{21}) & D_{11} + D_{22} + i (D_{12} - D_{21})
  \\[5pt]
  D_{11} + D_{22} - i (D_{12} - D_{21})  & D_{22} - D_{11} - i (D_{12} + D_{21})
 \end{pmatrix}\ .
\end{equation}
The various elements have phases with $\phi$ and/or $\psi$, i.e. have $m=\pm 1$. These of course cancel out in the parafermion
term in \eqn{s1c1},  leaving an operator with scaling dimension $\displaystyle 2+\nicefrac1k$ as advertized. Then, the $\b$-function becomes \eqn{betaFunclambda}. We stress that the appearance of the matrix $M$ is very important. For example, if this was the identity, then the dimension of the corresponding term 
would have been $\displaystyle 2-\frac{1}{2k}$, resulting to the negative $\b$-function  \cite{Itsios:2014lca} of the $\l$-deformed model based on the  $SO(3)_k/SO(2)$ coset CFT constructed in \cite{Sfetsos:2013wia}.

\section{Interpolating between exact CFTs}
\label{Sec:4}

In this section we will construct an integrable model interpolating between exact coset CFTs times a linear dilaton with the appropriate levels and  background charges. We will show that the target space of this integrable model interpolates  under the renormalization group flow, between two spacetimes of a cosmological and a black hole interpretation in the IR and UV fixed points, respectively.

\no
Let us start with a Minkowski signature background corresponding to the $\l$-deformed $SL(2,\IR)_{-k_1}\times SL(2,\IR)_{-k_2}/SL(2,\IR)_{-k_1-k_2}$  $\s$-model explicitly constructed in~\cite{Sfetsos:2014cea}. This model is classically integrable and it interpolates between the exact CFTs for $k_1 < k_2$, as shown below in Figure~\ref{RGflows.CFTs}.
\begin{figure}[H]
\centering
  \begin{tikzcd}
&\text{IR:}\quad SL(2,\IR)_{-k_1}\times SL(2,\IR)_{-k_2}/SL(2,\IR)_{-k_1-k_2} \arrow[d,dashrightarrow]\\
& \text{UV:}\quad SL(2,\IR)_{-k_1}\times SL(2,\IR)_{k_1-k_2}/SL(2,\IR)_{-k_2}
\end{tikzcd}
\caption{RG flows between exact CFTs associated to \eqref{action.levels}, \eqref{action.levels.drie} and \eqref{dilexact}.} 
\label{RGflows.CFTs}
\end{figure}
\no
Its action has zero antisymmetric tensor and conveniently can be written as
\be
\label{action.levels}
S=\frac{1}{\pi}\int \text{d}^2\s \left(-\text{e}^0_+\text{e}^0_-+\text{e}^1_+\text{e}^1_-+\text{e}^2_+\text{e}^2_-\right)\, ,
\ee
in terms of 
\be
\label{action.levels.drie}
\begin{split}
&\text{e}^0_\pm=\sqrt{\frac{(k_1+k_2)(1-\l)}{\L Z}}\left(\l_0^2\b_0\del_\pm\a_0+\l_0^{-2}\a_0\del_\pm\b_0-\del_\pm\g_0\right)\,,
\\
&\text{e}^1_\pm=\sqrt{\frac{k_1k_2}{k_1+k_2}\frac{Z}{\L(1-\l)}}\cos\frac{\psi}{2}
\left(\l_0 \sqrt{\b_0^2-1}\,\del_\pm \a_0-\l_0^{-1}\sqrt{\a_0^2-1}\,\del_\pm \b_0\right)\,,
\\
&\text{e}^2_\pm=\sqrt{\frac{k_1k_2}{k_1+k_2}\frac{Z}{\L(1-\l)}}\sin\frac{\psi}{2}
\left(\l_0 \sqrt{\b_0^2-1}\,\del_\pm \a_0+\l_0^{-1}\sqrt{\a_0^2-1}\,\del_\pm \b_0\right)\,,
\end{split}
\ee
where
\be
\label{hufi}
\begin{split}
&\L=(\a_0^2-1)(\b_0^2-1)-\g_0^2\, ,\quad \g_0=\sqrt{(\a_0^2-1)(\b_0^2-1)}\cos\psi\,,
 \\
&\l_0= \sqrt{k_1\ov k_2} \, ,\quad Z=8\l+(1-\l)(\l_0+\l_0^{-1})^2
\end{split}
\ee
and the dilaton equals
\be
\label{dilexact}
\text{e}^{-2\Phi}=\L\,.
\ee
The RG flow equation \eqn{RGflow} implies that
\begin{equation}
 \label{beta.coset}
 \frac{\text{d}\l}{\text{d}t} =\frac{2\l(1-\l_1^{-1}\l)(1-\l_2^{-1}\l)(1-\l^{-1}_3\l)}
{(k_1+k_2)(1-\l_f^{-1}\l)^2}\,,
\end{equation}
where
 \begin{equation}
\label{fixp}
\l_1=\frac{1+\l_0^2}{1-3\l_0^2}\,,\quad \l_2=\frac{1+\l_0^2}{\l_0^2-3}\,,\quad
 \l_3=\left(\frac{1+\l_0^2}{1-\l_0^2}\right)^2\,,\quad \l_f^{-1}=1-\frac{8\l_0^2}{(1+\l_0^2)^2}\,,
\end{equation}
provided that the diffeomorphism is given by \eqn{diff}. 
The above $\b$-function is just minus the one computed in \cite{Sfetsos:2017sep} for the $\l$-deformed model based on the 
coset $G_{k_1}\times G_{k_2}/G_{k_1+k_2}$, but now specialized to $G_k=SL(2,\IR)_{-k}$. The reason is that  the signs of $k_1$ and $k_2$ are flipped since the group here is non-compact and we require a Minkowski signature spacetime.

\no
Before analyzing the RG flow described by \eqref{beta.coset} we note  that the background \eqref{action.levels} and the $\b$-function \eqref{beta.coset}
are invariant under the action of the symmetry~\cite{Sfetsos:2017sep}
\be
\label{sym.coset}
\l\to \frac{1-\l_3^{-1}\l}{\l_3^{-1}-\l_f^{-1}\l}\,,\quad k_1\to-k_1\,,\quad k_2\to-k_2\,.
\ee
The fixed points of this symmetry are at $\l=1$ and $\l=\l_f$, when the background \eqref{action.levels} 
becomes also singular. 
In addition, the fixed points of the RG flow, i.e. $\l=0$ and $\l_{1,2,3}$
 in \eqn{fixp}, are mapped as $0\leftrightarrow \l_3$ and $\l_1\leftrightarrow  \l_2$.

\no
Let us now analyze the four fixed points $\l=(0,\l_{1,2,3})$ of the $\b$-function \eqref{beta.coset}. 
Taking $k_2>k_1$ the various points are ordered as follows
\be
\begin{split}
&-1<\l_2<0<1<\l_3<\l_1<\l_f\,,\quad\text{for}\quad 0<\l_0<\sqrt{2}-1\,, \\
&\l_f<-1<\l_2<0<1<\l_3<\l_1\,,\quad\text{for}\quad \sqrt{2}-1<\l_0<\frac{1}{\sqrt{3}}\,,\\
&\l_1<\l_f<-1<\l_2<0<1<\l_3\,,\quad\text{for}\quad \frac{1}{\sqrt{3}}<\l_0<1\, ,
\end{split}
\ee
where  $\l_0$ is defined in \eqn{hufi}.
The $\b$-function \eqref{beta.coset} describes a regular $\s$-model \eqref{action.levels} in the domain $\l_2\leqslant\l\leqslant0$.
To analyze the nature of the fixed points $\l=0$ and $\l=\l_2$ we expand the $\b$-function \eqref{beta.coset} and we keep only the linear term
\be
 \frac{\text{d}\l}{\text{d}t} \simeq \frac{2\l}{k_1+k_2}\leqslant0\quad\text{and}\quad
 \frac{\text{d}\l}{\text{d}t} \simeq -\frac{2(\l-\l_2)}{k_2-k_1}\leqslant0\,.
\ee
Thus, the fixed points $\l=0$ and $\l=\l_2$ correspond to an IR and a UV fixed point, respectively.
The underlying CFTs at the aforementioned points were analyzed in~\cite{Sfetsos:2017sep}  and are summarized in Figure \ref{RGflows.CFTs}.
The RG flow satisfies Zamolodchikov's c-theorem~\cite{Zamolodchikov:1986gt}, that is $c_\text{IR}< c_\text{UV}$ with
\be
\label{central.charges.cosetir}
\begin{split}
c_\text{IR}& ={3k_1\ov k_1-2} + {3k_2\ov k_2-2} - {3(k_1+k_2)\ov (k_1+k_2)-2}
\\
 & = 
3+ 6 \bigg({1\ov k_1}+\frac{1}{k_2}-{1\ov k_1+k_2}\bigg)+{\cal O}\left(\frac{1}{k_{1,2}^2}\right)\ 
\end{split}
\ee
and
\be
\begin{split}
\label{central.charges.cosetuv}
c_\text{UV} & ={3k_1\ov k_1-2} + {3(k_2-k_1)\ov (k_2-k_1)-2} - {3 k_2\ov k_2-2}
\\
& = 3 + 6 \bigg({1\ov k_1}-\frac{1}{k_2}+{1\ov k_2-k_1}\bigg)+{\cal O}\left(\frac{1}{k_{1,2}^2}\right)\ .
\end{split}
\ee

\subsection*{Taking the asymptotic limit}

To perform the asymptotic limit in \eqref{action.levels.drie} we first perform the transformation
\be
\label{limit.3d.levels}
\a_0=\text{e}^\a\,,\quad \b_0=\text{e}^\b\,,\quad \g_0=\text{e}^{\a+\b}\g\,,
\ee
which for large values of $\a$ and $\b$ yields a non-degenerate background such that 
the ratio $ \g=\nicefrac{\g_0}{\a_0\b_0}$ is kept finite.
The resulting action takes the form
\be
\label{action.levels.asym}
S=\frac{1}{\pi}\int \text{d}^2\s \left(-\frak{e}^0_+\frak{e}^0_-+\frak{e}^1_+\frak{e}^1_-+\frak{e}^2_+\frak{e}^2_-\right)\,,
\ee
with
\be
\label{action.levels.drie.assy}
\begin{split}
&\frak{e}^0_\pm=\sqrt{\frac{(k_1+k_2)(1-\l)}{Z(1-\g^2)}}\left((\l_0^2-\g)\del_\pm\a+(\l_0^{-2}-\g)\del_\pm\b-\del_\pm\g\right)\,,\\
&\frak{e}^1_\pm=\sqrt{\frac{k_1k_2}{k_1+k_2}\frac{Z}{2(1-\l)(1-\g)}}
\left(\l_0\,\del_\pm \a-\l_0^{-1}\,\del_\pm \b\right)\,,\\
&\frak{e}^2_\pm=\sqrt{\frac{k_1k_2}{k_1+k_2}\frac{Z}{2(1-\l)(1+\g)}}
\left(\l_0\,\del_\pm \a+\l_0^{-1}\,\del_\pm \b\right)\ 
\end{split}
\ee
where the target space of the action \eqref{action.levels.asym} has two shift isometries in the coordinates $\a$ and $\b$.
This is unlike \eqn{action.levels} which possesses no isometries.
In addition, the dilaton reads
\be
\label{dilexactasym}
\text{e}^{-2\Phi}=\text{e}^{2(\a+\b)}(1-\g^2)\, .
\ee
The background \eqref{action.levels.asym} is again invariant under the action of the symmetry~\eqref{sym.coset}.
The RG flow equation \eqref{RGflow} implies again \eqref{beta.coset}, with the diffeomorphism given by \eqn{diff}.
In what follows we will again consider $k_2>k_1$ and analyze the model \eqref{action.levels.asym} at
$\l=0$ and $\l=\l_2$, corresponding to the IR and UV fixed points respectively. 
To identify the IR CFT point at $\l=0$, we apply the following coordinate transformation
\be
\small
\a=\r+\ln\cosh t+\frac{k_2\,X}{k_1+k_2}\,,\quad
\b=-\r+\ln\cosh t+\frac{k_1\,X}{k_1+k_2}\,,\quad \g=\frac{2}{\cosh^2 t}-1\,,
\ee
yielding
\be
\label{IR.CFT.coset}
S_\text{IR}=\frac{1}{\pi}\int\text{d}^2\s\, \bigg((k_1+k_2)(-\del_+t\del_-t + \coth^2 t\,\del_+\r\del_-\r)
+\frac{k_1k_2}{k_1+k_2}\del_+X\del_-X\bigg)
\ee
and $\text{e}^{-2\Phi}=\text{e}^{2X}\sinh^2 t$. The two-dimensional curved part of the above metric has a cosmological 
interpretation \cite{Kounnas:1992wc}, it corresponds to the coset $SL(2,\IR)_{k_1+k_2}/U(1)$ CFT
and is a particular case of the more general cosmological model constructed in \cite{Nappi:1992kv}.
The corresponding central charge is
\be 
c_{\rm coset} ={3 (k_1+k_2)\ov (k_1+k_2)+2}-1= 2 - {6\ov k_1+k_2} + \dots\ ,  
\ee
to order ${\cal O}(\frac{1}{k_{1,2}})$.
Following the discussion at the end of Section~\ref{Interpolating.2d} we read
that the background charge of  the linear dilaton background corresponding to $X$ is $Q=-1$ and also $\a'=\nicefrac{1}{k_1}+\nicefrac{1}{k_2}$, for large levels.
Therefore the central charge is
\be
 c_{\ell.\text{d}}= 1 + 6\a'Q^2 = 1+ 6 \Big({1\ov k_1}+ {1\ov k_2}\Big)\  .
\ee
Hence $c_{\rm IR} = c_{\rm coset}+  c_{\ell.\text{d}}$ 
coincides with the large level expansion of the central charge of the IR theory in \eqn{central.charges.cosetir}.

\no
In the same spirit at the UV CFT point $\l=\l_2$, we apply the coordinate transformation
\be
\small
\a=t+\ln\cosh \rho+\frac{k_2\,X}{k_2-k_1}\,,\quad
\b=-t+\ln\cosh \rho+\frac{k_1\,X}{k_1-k_2}\,,\quad \g=1-\frac{2}{\cosh^2\rho}\,,
\ee
yielding 
\be
\label{UV.CFT.coset}
S_\text{UV}=\frac{1}{\pi}\int\text{d}^2\s\, \bigg((k_2-k_1)(\del_+\rho\del_-\rho-\coth^2\rho\,\del_+ t\del_- t)
+\frac{k_1k_2}{k_2-k_1}\del_+X\del_-X \bigg)
\ee
and $\text{e}^{-2\Phi}=\text{e}^{2X}\sinh^2\rho$. 
The two-dimensional curved part of the above metric has a black hole interpretation and corresponds to the coset CFT 
$SL(2,\IR)_{k_1-k_2}/U(1)$~\cite{wittenbh}. In particular it parametrizes the part of the black hole outside the horizon. 
The corresponding central charge to order ${\cal O}(\frac{1}{k_{1,2}})$ equals to
\be 
c_{\rm coset} ={3 (k_1-k_2)\ov (k_1-k_2)+2}-1= 2 + {6\ov k_2-k_1} + \dots\ ,  
\ee
The background charge of the linear dilaton corresponding to $X$ is $Q=-1$ and $\a'=\nicefrac{1}{k_1}-\nicefrac{1}{k_2}$, again for large levels. 
Therefore the central charge is
\be
c_{\ell.\text{d}}= 1 + 6\a'Q^2 = 1+ 6\Big({1\ov k_1}- {1\ov k_2}\Big)\  .
\ee
Hence $c_{\rm UV} = c_{\rm coset}+  c_{\ell.\text{d}}$ 
coincides with the large level expansion of the central charge of the UV theory in \eqn{central.charges.cosetuv}. 

\no
 To put it concisely, in the asymptotic limit \eqref{limit.3d.levels} for large values of $\a$ and $\b$, the integrable model \eqref{action.levels.asym} interpolates between exact CFTs times a linear dilaton for $k_1 < k_2$, as shown below in Figure \ref{RGflows.CFTs.asym}. The corresponding target spacetimes
have a cosmological \eqref{IR.CFT.coset} and black hole \eqref{UV.CFT.coset} interpretation in the IR and UV fixed points, respectively.
\begin{figure}[H]
\centering
  \begin{tikzcd}
&\text{IR:}\quad SL(2,\IR)_{k_1+k_2}/U(1) \;\; \times \;\;  \mathbb{R}_{Q} \arrow[d,dashrightarrow]\\
& \text{UV:}\quad SL(2,\IR)_{k_1-k_2}/U(1) \;\; \times \;\;  \mathbb{R}_{Q}
\end{tikzcd}
\caption{RG flows associated to \eqref{action.levels.asym}, \eqref{action.levels.drie.assy} and \eqref{dilexactasym}. The IR and UV fixed points are described in terms of \eqref{IR.CFT.coset} and \eqref{UV.CFT.coset}, respectively.} 
\label{RGflows.CFTs.asym}
\end{figure}

\section{ Concluding remarks}
\label{Sec:5}

 In the present work we constructed and studied a new class of integrable $\s$-models which interpolate between exact CFTs involving 
compact groups and a linear dilaton background  in the IR and hyperbolic spaces in the UV. In the prototype example,  the CFT IR theory is just 
the Euclidean space including a linear dilaton background and that in the UV the $H_2$ space. We showed that this model arises in the asymptotic limit of the $\l$-deformed $SL(2,\IR)_{-k}/SO(1,1)$ $\s$-model.  Armed with this observation we provided a way to construct new integrable 
$\s$-models by taking the asymptotic limit of certain $\l$-deformed models. In that spirit, we explicitly worked out the details starting with
the $\l$-deformed $SO(3,1)_{-k}/SO(3)_{-k}$ $\s$-model. We find that the latter in the asymptotic limit interpolates between the coset CFT $SO(3)_k/SO(2)$ times a linear dilaton background in the IR and the $H_3$ space in the UV. We have noted that the procedure to discover the UV theory involves a zoom-in limit which is 
quite distinct from the non-Abelian T-dual limit of $\l$-deformed $\s$-models as this was defined in~\cite{Sfetsos:2013wia}. 
In all cases considered we have uncovered the CFT nature of the operators driving the perturbation away from the conformal points. 
Exploring the group theoretical structure underlying this new limiting procedure for general $\l$-deformations based on coset 
CFTs certainly is worth investigating. We expect that the resulting backgrounds will be rather 
simple and will have isometries. Specifically, for the model obtained by taking the asymptotic 
limit of the $d$-dimensional $\l$-deformed model based on the CFT coset 
$SO(d,1)_{-k}/SO(d)_{-k}$,  we expect $d-1$ Abelian isometries.

\no
In addition, we have constructed  a model having a conformal fixed point in the UV aside from the one in the IR. 
To obtain its action we considered the asymptotic limit of the $\l$-deformed $SL(2,\IR)_{-k_1}\times SL(2,\IR)_{-k_2}/SL(2,\IR)_{-k_1-k_2}$ $\s$-model. 
The model we found interpolates under the renormalization group flow, between exact coset
CFTs times a linear dilaton background, with a cosmological and black hole interpretation
in the IR and UV, respectively. This is quite remarkable since at intermediate energy
scales the target spacetime does not have such a simple interpretation.
The RG flow from a cosmological towards a black hole spacetime appears to be acceptable from a physical
point of view. One expects that at low energies (IR) the physical degrees of freedom
can be accommodated easier in a spacetime that it is being created whereas at high
energies (UV) we can fit more degrees of freedom inside the black hole horizon. Needless
to say that these statements should be taken with caution and further investigations
should be undertaken.

\no
The results of the present work can be extended for higher dimensional cases and for different non-compact groups, 
by considering asymptotic limits of the corresponding $\l$-deformed models. We expect a richer structure if one starts with the two-parameter integrable models constructed in \cite{Georgiou:2017jfi} and the multi-parameter ones in~\cite{Georgiou:2018hpd,Georgiou:2018gpe}. 

\no
Another extension could involve discovering and studying other kinds of asymptotic limits, especially those taken in $\s$-models with Lorentzian target spaces. Such type of limits was considered in  \cite{Hoare:2015wia} for the background corresponding to the four-dimensional  coset $SO(3,2)_{-k}/SO(3,1)_{-k}$ CFT \cite{Bars:1991zt}
 (in the global coordinate system of \cite{Bars:1992sr}). In the three-dimensional model \eqn{met3d}, \eqref{3ddil} this limit corresponds in taking 
 $u,v\to\infty$ and keep $\nicefrac{u}{v}$ finite. It would be worth investigating this limit away from the conformal point.

\no
A future direction, worth considering is to embed the interpolating solutions in type-II supergravity with non-trivial Ramond--Ramond fluxes. Progress in this direction has already been undertaken in~\cite{Sfetsos:2014cea,Demulder:2015lva,Itsios:2019izt,Chervonyi:2016ajp,Borsato:2016ose} and it is based on $\l$-deformations for group and coset spaces~\cite{Sfetsos:2013wia}. Note that such embeddings are generically non-super\-symmetric and their stability analysis needs to be tested separately~\cite{Itsios:2021xwh}.

\no
Another potential extension involves the study of analogue asymptotic limits for the Yang--Baxter deformations of principal chiral models in symmetric coset spaces constructed in~\cite{Delduc:2013fga,Delduc:2013qra}. This class of deformations is related to the $\l$-deformed ones via Poisson--Lie T-duality and an analytic continuation of the coordinates and parameters of the $\s$-model~\cite{Vicedo:2015pna, Hoare:2015gda,Sfetsos:2015nya,Klimcik:2015gba,Klimcik:2016rov}. 
It would be worth studying analogue asymptotic limits in the aforementioned deformations and whether they are related to the maximal boost limit considered in~\cite{Hoare:2016hwh}. In this case the integrable interpolations are expected to be different due to the lack of a conformal point.

\subsection*{Acknowledgements}

We would like to thank P.M. Petropoulos for carefully reading of the manuscript and useful suggestions.\\
The research work of K.~Sfetsos was supported by the Hellenic Foundation for
Research and Innovation (H.F.R.I.) under the ``First Call for H.F.R.I.
Research Projects to support Faculty members and Researchers and
the procurement of high-cost research equipment grant'' (MIS 1857, Project Number: 16519).\\
The research work of K.~Siampos has received funding from the Hellenic Foundation
for Research and Innovation (H.F.R.I.) and the General Secretariat for Research and Technology (G.S.R.T.),
under grant agreement No 15425.



\begin{thebibliography}{99}



\bibitem{coset}
  P.~Goddard, A.~Kent and D.~I.~Olive,
  {\it Virasoro algebras and coset space models},\\
  \href{https://www.sciencedirect.com/science/article/abs/pii/0370269385911451}{Phys. Lett. {\bf B152} (1985), 88-92}.

\bibitem{gwzw}
  K.~Bardakci, E.~Rabinovici and B.~Saering,
  {\it String models with c < 1 components},
  \href{https://www.sciencedirect.com/science/article/abs/pii/0550321388904701}{Nucl. Phys. {\bf B299} (1988), 151-182};\hfill\break
  H.~J.~Schnitzer,
 {\it A path integral construction of superconformal field theories from a gauged supersymmetric Wess-Zumino-Witten action},\\
  \href{https://www.sciencedirect.com/science/article/abs/pii/0550321389904732}{Nucl. Phys. {\bf B324} (1989), 412-426};\hfill\break
  D.~Karabali, Q.~H.~Park, H.~J.~Schnitzer and Z.~Yang,
  {\it A GKO construction based on a path integral formulation of gauged Wess-Zumino-Witten actions},\\
  \href{https://www.sciencedirect.com/science/article/abs/pii/0370269389911209}{Phys. Lett. {\bf B216} (1989), 307-312}.
  

\bibitem{wittenbh}
  E.~Witten,
  {\it On string theory and black holes},
  \href{https://journals.aps.org/prd/abstract/10.1103/PhysRevD.44.314}{Phys. Rev. {\bf D44} (1991), 314-324}.
  
  \bibitem{Bars:1991pt}
I.~Bars and K.~Sfetsos,
{\it Generalized duality and singular strings in higher dimensions},
\href{https://www.worldscientific.com/doi/abs/10.1142/S0217732392000963}{Mod. Phys. Lett. \textbf{A7} (1992), 1091-1104},
[\href{https://arxiv.org/abs/hep-th/9110054}{\ttfamily hep-th/9110054}].

\bibitem{Fradkin:1991ie}
E.~S.~Fradkin and V.~Y.~Linetsky,
{\it On space-time interpretation of the coset models in D \ensuremath{<} 26 critical string theory},
\href{https://www.sciencedirect.com/science/article/abs/pii/0370269392909598}{Phys. Lett. \textbf{B277} (1992), 73-78}.


\bibitem{Dijkgraaf:1991ba}
R.~Dijkgraaf, H.~L.~Verlinde and E.~P.~Verlinde,
{\it String propagation in a black hole geometry},
\href{https://www.sciencedirect.com/science/article/pii/0550321392902376}{Nucl. Phys.  \textbf{B371} (1992), 269-314}.

\bibitem{Bars:1991zt}
I.~Bars and K.~Sfetsos,
{\it A Superstring theory in four curved space-time dimensions},
\href{https://www.sciencedirect.com/science/article/pii/037026939290746Q?via\%3Dihub}{Phys. Lett.  \textbf{B277} (1992), 269-276},
[\href{https://arxiv.org/abs/hep-th/9111040}{\ttfamily hep-th/9111040}].

\bibitem{Horne:1991gn}
J.~H.~Horne and G.~T.~Horowitz,
{\it Exact black string solutions in three-dimensions},
\href{https://www.sciencedirect.com/science/article/pii/055032139290536K?via\%3Dihub}{Nucl. Phys.  \textbf{B368} (1992), 444-462},
[\href{https://arxiv.org/abs/hep-th/9108001}{\ttfamily hep-th/9108001}].
  
 

\bibitem{Forgacs:1989ac}
P.~Forgacs, A.~Wipf, J.~Balog, L.~Feher and L.~O'Raifeartaigh,
{\it Liouville and Toda Theories as Conformally Reduced WZNW Theories},
\href{https://www.sciencedirect.com/science/article/pii/S0370269389800255?via\%3Dihub}{Phys. Lett. \textbf{B227} (1989), 214-220}.

\bibitem{Petropoulos:1989fc}
P.~M.~S.~Petropoulos,
{\it Comments on $SU(1,1)$ string theory},\\
\href{https://www.sciencedirect.com/science/article/pii/037026939090819R?via\%3Dihub}{Phys. Lett. \textbf{B236} (1990), 151-158}.


 \bibitem{Bars:1989ph}
I.~Bars,
{\it Heterotic Superstring Vacua in 4-$D$ Based on Noncompact Affine Current Algebras},
\href{https://www.sciencedirect.com/science/article/pii/0550321390906592?via\%3Dihub}{Nucl. Phys. \textbf{B334} (1990), 125-171}.


 
 
\bibitem{Petropoulos:2006py}
P.~M.~Petropoulos and K.~Sfetsos,
{\it Non-Abelian coset string backgrounds from asymptotic and initial data},
\href{https://iopscience.iop.org/article/10.1088/1126-6708/2007/04/033}{JHEP \textbf{04} (2007), 033},
[\href{https://arxiv.org/abs/hep-th/0610055}{\ttfamily hep-th/0610055}].

\bibitem{Sfetsos:2013wia}
  K.~Sfetsos,
    {\it Integrable interpolations: From exact CFTs to non-Abelian T-duals},\hfill\break
  \href{https://www.sciencedirect.com/science/article/pii/S0550321314000066?via\%3Dihub}{Nucl. Phys. {\bf B880} (2014), 225-246},
 [\href{http://arxiv.org/abs/1312.4560}{\ttfamily 1312.4560}].
 
  \bibitem{Hollowood:2014rla}
  T.~J.~Hollowood, J.~L.~Miramontes and D.~M.~Schmidtt,
  {\it Integrable deformations of strings on symmetric spaces},
  \href{https://link.springer.com/article/10.1007\%2FJHEP11\%282014\%29009}{JHEP {\bf 1411} (2014), 009},
 [\href{http://arxiv.org/abs/1407.2840}{\ttfamily 1407.2840}].
 
 \bibitem{Demulder:2015lva}
S.~Demulder, K.~Sfetsos and D.~C.~Thompson,
{\it Integrable $\lambda$-deformations: Squashing Coset CFTs and $AdS_5\times S^5$},
\href{https://link.springer.com/article/10.1007\%2FJHEP07\%282015\%29019}{JHEP \textbf{07} (2015), 019},
[\href{https://arxiv.org/abs/1504.02781}{\ttfamily 1504.02781}].


\bibitem{Itsios:2014lca}
G.~Itsios, K.~Sfetsos and K.~Siampos,
{\it The all-loop non-Abelian Thirring model and its RG flow},
\href{https://www.sciencedirect.com/science/article/pii/S0370269314003037?via\%3Dihub}{Phys. Lett. \textbf{B733} (2014), 265-269},
[\href{https://arxiv.org/abs/1404.3748}{\ttfamily 1404.3748}].


\bibitem{Sfetsos:2014jfa}
K.~Sfetsos and K.~Siampos,
{\it Gauged WZW-type theories and the all-loop anisotropic non-Abelian Thirring model},
\href{https://www.sciencedirect.com/science/article/pii/S0550321314001953?via\%3Dihub}{Nucl. Phys. \textbf{B885} (2014), 583-599},
[\href{https://arxiv.org/abs/1405.7803}{1405.7803}].

\bibitem{Appadu:2015nfa}
C.~Appadu and T.~J.~Hollowood,
{\it Beta function of k deformed $AdS_5 \times S^5$ string theory},
\href{https://link.springer.com/article/10.1007\%2FJHEP11\%282015\%29095}{JHEP \textbf{11} (2015), 095},
[\href{https://arxiv.org/abs/1507.05420}{\ttfamily 1507.05420}].

 \bibitem{Sfetsos:2014cea}
  K.~Sfetsos and D.~C.~Thompson,
  {\it Spacetimes for $\l$-deformations},\hfill\break
  \href{https://link.springer.com/article/10.1007/JHEP12(2014)164}{JHEP {\bf 1412} (2014), 164},
[\href{http://arxiv.org/abs/arXiv:1410.1886}{\ttfamily 1410.1886}].

  \bibitem{Sfetsos:2017sep}
K.~Sfetsos and K.~Siampos,
{\it Integrable deformations of the $G_{k_1} \times G_{k_2}/G_{k_1+k_2}$ coset CFTs},
\href{https://www.sciencedirect.com/science/article/pii/S0550321317303966?via\%3Dihub}{Nucl. Phys. \textbf{B927} (2018), 124-139},
[\href{https://arxiv.org/abs/1710.02515}{\ttfamily 1710.02515}].

\bibitem{Ecker}
G.~Ecker and J.~Honerkamp,
{\it Application of invariant renormalization to the nonlinear chiral invariant
pion Lagrangian in the one-loop approximation},\hfill\break
\href{http://www.sciencedirect.com/science/article/pii/0550321371904688}{Nucl. Phys. {\bf B35} (1971), 481-492}.\hfill\break
J.~Honerkamp,
{\it Chiral multiloops},
\href{http://www.sciencedirect.com/science/article/pii/0550321372902994}{Nucl. Phys. {\bf B36} (1972), 130-140}.

\bibitem{Friedan:1980jf}
 D.~Friedan,
 {\it Nonlinear models in two epsilon dimensions},\hfill\break
 \href{http://journals.aps.org/prl/abstract/10.1103/PhysRevLett.45.1057}{Phys. Rev. Lett. {\bf 45} (1980), 1057-1060}
and {\it Nonlinear models in $2+\varepsilon$ dimensions},\hfill\break
 \href{http://www.sciencedirect.com/science/article/pii/0003491685903847}{Annals Phys.\  {\bf 163} (1985), 318-419}.

 
 \bibitem{Callan:1985ia}
C.~G.~Callan, E.~J.~Martinec, M.~J.~Perry and D.~Friedan,
{\it Strings in Background Fields},
\href{https://www.sciencedirect.com/science/article/pii/0550321385905061?via\%3Dihub}{Nucl. Phys. \textbf{B262} (1985), 593-609}.



\bibitem{Sklyanin:1980ij}
  E.~K.~Sklyanin,
  {\it Quantum version of the method of inverse scattering problem},\\
\href{http://link.springer.com/article/10.1007\%2FBF01091462} {J. Sov. Math. {\bf 19} (1982) 1546},
   Zap.\ Nauchn.\ Semin.\  {\bf 95} (1980) 55.
  
  \bibitem{Maillet:1985ek}
  J.~M.~Maillet,
 {\it New Integrable Canonical Structures in Two-dimensional Models},\\
 \href{http://www.sciencedirect.com/science/article/pii/0550321386903652}{Nucl. Phys. {\bf B269} (1986) 54}.

\bibitem{Maillet:1985ec}
  J.~M.~Maillet,
  {\it Hamiltonian Structures for Integrable Classical Theories From Graded Kac-moody Algebras},
 \href{http://www.sciencedirect.com/science/article/pii/037026938691289X}{Phys. Lett. {\bf B167} (1986) 401}.
 
  \bibitem{Hollowood:2015dpa}
T.~J.~Hollowood, J.~L.~Miramontes and D.~M.~Schmidtt,
{\it S-Matrices and Quantum Group Symmetry of k-Deformed Sigma Models},\\
\href{https://iopscience.iop.org/article/10.1088/1751-8113/49/46/465201}{J. Phys. A \textbf{49} (2016) no.46, 465201},
[\href{https://arxiv.org/abs/1506.06601}{\ttfamily 1506.06601}].
 
 \bibitem{Georgiou:2020bpx}
G.~Georgiou, K.~Sfetsos and K.~Siampos,
{\it A free field perspective of \ensuremath{\lambda}-deformed coset CFT\textquoteright{}s},
\href{https://link.springer.com/article/10.1007/JHEP07(2020)187}{JHEP \textbf{07} (2020), 187},
[\href{https://arxiv.org/abs/2004.10216}{\ttfamily 2004.10216}].
 
 \bibitem{Bars:1992ti}
I.~Bars and K.~Sfetsos,
{\it Global analysis of new gravitational singularities in string and particle theories},
\href{https://journals.aps.org/prd/abstract/10.1103/PhysRevD.46.4495}{Phys. Rev. \textbf{D46} (1992), 4495-4509},
[\href{https://arxiv.org/abs/hep-th/9205037}{\ttfamily hep-th/9205037}].

  \bibitem{Lykken:1988ut}
J.~D.~Lykken,
{\it Finitely reducible realizations of the $N=2$ superconformal algebra},\\
\href{https://www.sciencedirect.com/science/article/pii/0550321389903295?via\%3Dihub}{Nucl. Phys. \textbf{B313} (1989), 473-491}. 
 


\bibitem{Fateev:1985mm}
V.~A. Fateev and A.~B.~Zamolodchikov, {\it Parafermionic currents in the
  two-dimensional conformal quantum field theory and selfdual critical points
  in Z(n) invariant statistical systems}, \href{http://www.jetp.ac.ru/cgi-bin/e/index/e/62/2/p215?a=list}{Sov. Phys. JETP {\bfseries
  62} (1985), 215-225}.
  
\bibitem{Georgiou:2017aei}
G.~Georgiou, E.~Sagkrioti, K.~Sfetsos and K.~Siampos,
{\it Quantum aspects of doubly deformed CFTs},
\href{https://www.sciencedirect.com/science/article/pii/S0550321317301256?via\%3Dihub}{Nucl. Phys. \textbf{B919} (2017), 504-522},
[\href{https://arxiv.org/abs/1703.00462}{\ttfamily 1703.00462}].


  \bibitem{Georgiou:2017jfi}
  G.~Georgiou and K.~Sfetsos,
  {\it Integrable flows between exact CFTs},\\
  \href{https://link.springer.com/article/10.1007\%2FJHEP11\%282017\%29078}{JHEP {\bf 1711}  (2017), 078},
   [\href{https://arxiv.org/abs/1707.05149}{\ttfamily 1707.05149}].
  
 \bibitem{Bardacki:1990wj}
K.~Bardakci, M.~J.~Crescimanno and E.~Rabinovici, 
{\it Parafermions from coset models},
  \href{https://doi.org/10.1016/0550-3213(90)90365-K}{Nucl. Phys. {\bfseries B344} (1990), 344-370}.


\bibitem{Zamolodchikov:1986gt}
  A.~B.~Zamolodchikov,
 {\it Irreversibility of the flux of the renormalization group in a 2D field theory},
\href{http://jetpletters.ru/ps/1413/article_21504.shtml}{JETP Lett.  {\bf 43} (1986), 730-732}.

\bibitem{Kounnas:1992wc}
C.~Kounnas and D.~L\"ust,
{\it Cosmological string backgrounds from gauged WZW models},
\href{https://www.sciencedirect.com/science/article/abs/pii/037026939291361C?via\%3Dihub}{Phys. Lett. \textbf{B289} (1992), 56-60},
[\href{https://arxiv.org/abs/hep-th/9205046}{\ttfamily hep-th/9205046}].

\bibitem{Nappi:1992kv}
C.~R.~Nappi and E.~Witten,
{\it A closed, expanding universe in string theory},\\
\href{https://www.sciencedirect.com/science/article/pii/037026939290888B?via\%3Dihub}{Phys. Lett. \textbf{B293} (1992), 309-314},
[\href{https://arxiv.org/abs/hep-th/9206078}{\ttfamily hep-th/9206078}].

\bibitem{Georgiou:2018hpd}
G.~Georgiou and K.~Sfetsos,
{\it Novel all loop actions of interacting CFTs: Construction, integrability and RG flows},
\href{https://www.sciencedirect.com/science/article/pii/S0550321318302992?via\%3Dihub}{Nucl. Phys. \textbf{B937} (2018), 371-393},
[\href{https://arxiv.org/abs/1809.03522}{\ttfamily 1809.03522}].

\bibitem{Georgiou:2018gpe}
G.~Georgiou and K.~Sfetsos,
{\it The most general $\lambda$-deformation of CFTs and integrability},
\href{https://link.springer.com/article/10.1007\%2FJHEP03\%282019\%29094}{JHEP \textbf{03} (2019), 094},
[\href{https://arxiv.org/abs/1812.04033}{\ttfamily 1812.04033}].

\bibitem{Hoare:2015wia}
B.~Hoare and A.~A.~Tseytlin,
{\it Type IIB supergravity solution for the T-dual of the $\eta$-deformed AdS$_{5} \times$ S$^{5}$ superstring},
\href{https://link.springer.com/article/10.1007\%2FJHEP10\%282015\%29060}{JHEP \textbf{10} (2015), 060},
[\href{https://arxiv.org/abs/1508.01150}{\ttfamily 1508.01150}].

\bibitem{Bars:1992sr}
I.~Bars and K.~Sfetsos,
{\it Conformally exact metric and dilaton in string theory on curved space-time},
\href{https://journals.aps.org/prd/abstract/10.1103/PhysRevD.46.4510}{Phys. Rev. D \textbf{D46} (1992), 4510-4519},
[\href{https://arxiv.org/abs/hep-th/9206006}{\ttfamily hep-th/9206006}].


\bibitem{Itsios:2019izt}
G.~Itsios and K.~Sfetsos,
{\it $AdS$ solutions and $\lambda$-deformations},\\
\href{https://www.sciencedirect.com/science/article/pii/S0550321320300468?via\%3Dihub}{Nucl. Phys. \textbf{B953} (2020), 114960},
[\href{https://arxiv.org/abs/1911.12371}{\ttfamily 1911.12371}].

\bibitem{Chervonyi:2016ajp}
Y.~Chervonyi and O.~Lunin,
{\it Supergravity background of the $\lambda$-deformed AdS$_3 \times$ S$^3$ supercoset},
\href{https://www.sciencedirect.com/science/article/pii/S0550321316302061?via\%3Dihub}{Nucl. Phys. \textbf{B910} (2016), 685-711},
[\href{https://arxiv.org/abs/1606.00394}{1606.00394}].

\bibitem{Borsato:2016ose}
R.~Borsato and L.~Wulff,
{\it Target space supergeometry of $\eta$ and $\lambda$-deformed strings},
\href{https://link.springer.com/article/10.1007/JHEP10(2016)045}{JHEP \textbf{10} (2016), 045},
[\href{https://arxiv.org/abs/1608.03570}{1608.03570}].


\bibitem{Itsios:2021xwh}
G.~Itsios, P.~Panopoulos, K.~Sfetsos and D.~Zoakos,
{\it On the stability of $AdS$ backgrounds with $\lambda$-deformed factors},
\href{https://link.springer.com/article/10.1007\%2FJHEP07\%282021\%29054}{JHEP \textbf{07} (2021), 054},
[\href{https://arxiv.org/abs/2103.12761}{\ttfamily 2103.12761}].

\bibitem{Delduc:2013fga}
F.~Delduc, M.~Magro and B.~Vicedo,
{\it On classical $q$-deformations of integrable sigma-models},
\href{https://link.springer.com/article/10.1007/JHEP11(2013)192}{JHEP \textbf{11} (2013), 192},
[\href{https://arxiv.org/abs/1308.3581}{\ttfamily 1308.3581}].

\bibitem{Delduc:2013qra}
F.~Delduc, M.~Magro and B.~Vicedo,
{\it An integrable deformation of the $AdS_5 \times S^5$ superstring action},
\href{https://journals.aps.org/prl/abstract/10.1103/PhysRevLett.112.051601}{Phys. Rev. Lett. {\bf 112} (2014), 051601},
[\href{https://arxiv.org/abs/1309.5850}{\ttfamily 1309.5850}].
  
   \bibitem{Vicedo:2015pna}
  B.~Vicedo,
  {\it Deformed integrable $\sigma$-models, classical $R$-matrices and classical exchange algebra on Drinfel'd doubles},
  \href{https://iopscience.iop.org/article/10.1088/1751-8113/48/35/355203}{J. Phys. A: Math. Theor. {\bf 48} (2015), 355203},
 [\href{http://arxiv.org/abs/1504.06303}{\ttfamily 1504.06303}].

\bibitem{Hoare:2015gda}
  B.~Hoare and A.~A.~Tseytlin,
  {\it On integrable deformations of superstring sigma models related to $AdS_n \times S^n$ supercosets},
  \href{https://www.sciencedirect.com/science/article/pii/S0550321315002035?via\%3Dihub}{Nucl.\ Phys.\ {\bf B897} (2015), 448-478},
    [\href{http://arxiv.org/abs/1504.07213}{\ttfamily 1504.07213}].
    

 \bibitem{Sfetsos:2015nya}
  K.~Sfetsos, K.~Siampos and D.~C.~Thompson,
 {\it Generalised integrable $\lambda$- and $\eta$-deformations and their relation},
  \href{https://linkinghub.elsevier.com/retrieve/pii/S0550321315003004}{Nucl. Phys. {\bf B899} (2015), 489-512},
  [\href{http://arxiv.org/abs/1506.05784}{\ttfamily 1506.05784}].

\bibitem{Klimcik:2015gba}
C. Klim\v c\'\i k,
  {\it $\eta$ and $\lambda$ deformations as ${\cal E}$-models}, \hfill\break
\href{https://www.sciencedirect.com/science/article/pii/S0550321315003302?via\%3Dihub}{Nucl.\ Phys.\ {\bf B900} (2015), 259-272},
  [\href{http://arxiv.org/abs/1508.05832}{\ttfamily 1508.05832}].


\bibitem{Klimcik:2016rov}
C.~Klim\v{c}\'\i{}k,
{\it Poisson\textendash{}Lie T-duals of the bi-Yang\textendash{}Baxter models},\\
\href{https://www.sciencedirect.com/science/article/pii/S0370269316303380?via\%3Dihub}{Phys. Lett. \textbf{B760} (2016), 345-349},
[\href{https://arxiv.org/abs/1606.03016}{\ttfamily 1606.03016}].


\bibitem{Hoare:2016hwh}
B.~Hoare and S.~J.~van Tongeren,
{\it On jordanian deformations of AdS$_5$ and supergravity},
\href{https://iopscience.iop.org/article/10.1088/1751-8113/49/43/434006}{J. Phys. \textbf{A49} (2016) no.43, 434006},
[\href{https://arxiv.org/abs/1605.03554}{\ttfamily 1605.03554}].





\end{thebibliography}

\end{document}